\documentclass[aps,prb,twocolumn,superscriptaddress,showpacs]{revtex4}

\usepackage{graphicx}
\usepackage{dcolumn}
\usepackage{bm}

\begin{document}

\title{Thermal conductivity of IPA-CuCl$_3$: Evidences of ballistic magnon transport and limited applicability of the Bose-Einstein condensation model}

\author{Z. Y. Zhao}
\thanks{Present Address: Department of Physics and Astronomy, University of Tennessee, Knoxville, Tennessee 37996-1200, USA}

\affiliation{Hefei National Laboratory for Physical Sciences at Microscale, University of Science and Technology of China, Hefei, Anhui 230026, People's Republic of China}

\author{B. Tong}
\affiliation{Hefei National Laboratory for Physical Sciences at Microscale, University of Science and Technology of China, Hefei, Anhui 230026, People's Republic of China}

\author{X. Zhao}
\email{xiazhao@ustc.edu.cn}

\affiliation{School of Physical Sciences, University of Science and Technology of China, Hefei, Anhui 230026, People's Republic of China}

\author{L. M. Chen}
\affiliation{College of Electronic Science Engineering, Nanjing University of Post and Telecommunication, Nanjing, Jiangsu 211106, People's Republic of
China}

\author{J. Shi}
\affiliation{Department of Physics, University of Science and Technology of China, Hefei, Anhui 230026, People's Republic of China}

\author{F. B. Zhang}
\affiliation{Hefei National Laboratory for Physical Sciences at Microscale, University of Science and Technology of China, Hefei, Anhui 230026, People's Republic of China}

\author{J. D. Song}
\affiliation{Hefei National Laboratory for Physical Sciences at Microscale, University of Science and Technology of China, Hefei, Anhui 230026, People's Republic of China}

\author{S. J. Li}
\affiliation{Hefei National Laboratory for Physical Sciences at Microscale, University of Science and Technology of China, Hefei, Anhui 230026, People's Republic of China}

\author{J. C. Wu}
\affiliation{Hefei National Laboratory for Physical Sciences at Microscale, University of Science and Technology of China, Hefei, Anhui 230026, People's Republic of China}

\author{H. S. Xu}
\affiliation{Hefei National Laboratory for Physical Sciences at Microscale, University of Science and Technology of China, Hefei, Anhui 230026, People's Republic of China}

\author{X. G. Liu}
\affiliation{Hefei National Laboratory for Physical Sciences at Microscale, University of Science and Technology of China, Hefei, Anhui 230026, People's Republic of China}

\author{X. F. Sun}
\email{xfsun@ustc.edu.cn}

\affiliation{Hefei National Laboratory for Physical Sciences at Microscale, University of Science and Technology of China, Hefei, Anhui 230026, People's Republic of China}

\affiliation{Key Laboratory of Strongly-Coupled Quantum Matter Physics, Chinese Academy of Sciences, Hefei, Anhui 230026, People's Republic of China}

\affiliation{Collaborative Innovation Center of Advanced Microstructures, Nanjing, Jiangsu 210093, People's Republic of China}

\date{\today}

\begin{abstract}

The heat transport of the spin-gapped material (CH$_3$)$_2$CHNH$_3$CuCl$_3$ (IPA-CuCl$_3$), a candidate quantum magnet with Bose-Einstein condensation (BEC), is studied at ultra-low temperatures and in high magnetic fields. Due to the presence of the spin gap, the zero-field thermal conductivity ($\kappa$) is purely phononic and shows a ballistic behavior at $T <$ 1 K. When the gap is closed by magnetic field at $H = H_{c1}$, where a long-range antiferromanetic (AF) order of Cu$^{2+}$ moments is developed, the magnons contribute significantly to heat transport and exhibit a ballistic $T^3$ behavior at $T <$ 600 mK. In addition, the low-$T$ $\kappa(H)$ isotherms show sharp peaks at $H_{c1}$, which indicates a gap re-opening in the AF state ($H > H_{c1}$) and demonstrates limited applicability of the BEC model to IPA-CuCl$_3$.

\end{abstract}

\pacs{66.70.-f, 75.47.-m, 75.50.-y}

\maketitle

\section{INTRODUCTION}

Low-temperature thermal conductivity ($\kappa$) can probe the transport properties of various elementary excitations in solids.\cite{Berman, Ziman, Ashcroft} At very low temperatures, the phonon transport is known to exhibit a ballistic behavior since all the microscopic scatterings are smeared out and phonons are only scattered by the sample surface or boundary. This is the so-called boundary scattering limit and the $\kappa$ shows a simple $T^3$ dependence.\cite{Berman, Ziman, Ashcroft} The magnon excitations of an antiferromagnetically ordered material are known to have the same statistic law as the phonons. In addition, since the low-energy antiferromagnetic (AF) magnons also have linear dispersion, they are able to show the same $T^3$ dependence of thermal conductivity at very low temperatures.\cite{Berman, Ziman, Ashcroft} However, this ballistic transport of AF magnons has been rarely observed.\cite{Li_NCO} One reason is that the low-energy magnons and phonons can easily couple to each other.\cite{Wang_HMO, Zhao_GFO, Zhao_DFO} More serious is that the acoustic magnons are almost always gapped in the real antiferromagnets. In this regard, it is possible to switch on the magnon heat transport by applying magnetic field to close the anisotropy gap.\cite{Spin_flop, Li_NCO, Jin_NCO}. For example, in the AF insulator Nd$_2$CuO$_4$, a 10 T-field-induced increase of $\kappa$ was found to show a $T^3$ behavior at ultra-low temperatures, which was discussed to be the first observation of the AF magnon ballistic transport.\cite{Li_NCO} However, this explanation is not well grounded since the anisotropy gap can be closed only at the spin re-orientation field (much lower than 10 T for Nd$_2$CuO$_4$) and it will be re-opened at higher field.\cite{Spin_flop, Zhao_NCO}

It might be easier to probe the ballistic transport of the AF magnons in those quantum magnets that can exhibit the Bose-Einstein condensation (BEC).\cite{Review-1, Review-2, BEC-1, BEC-2, BEC-3, BEC-4, BEC-5, BEC-6, BEC-7} BEC denotes a collective occupation of bosons to the lowest single-particle state when temperature approaches zero. In some spin-gapped quantum magnets, the XY-type AF state induced by the magnetic field that closes the spin gap can be described as a BEC state. The spin Hamiltonian of these quantum magnets must contain the $U(1)$ symmetry, which requires a continuous uniaxial symmetry. In field-induced BEC state, the $U(1)$ symmetry spontaneously gets broken and, as a consequence, a gapless Goldstone mode is acquired. If this kind of low-energy AF magnons are not strongly coupled with phonons, they are likely to contribute to transporting heat.

Probing the magnon heat transport of BEC materials have been tried for the organic compound NiCl$_2$-4SC(NH$_2$)$_2$ (DTN), an $S =$ 1 chain system,\cite{Sun_DTN, Kohama} and the oxide Ba$_3$Mn$_2$O$_8$.\cite{Ke_BMO} It has been found that along the chain direction of DTN, the $\kappa$ shows a distinct enhancement when increasing field across the BEC phase boundary.\cite{Sun_DTN, Kohama} However, the reported data could not separate the magnon thermal conductivity from the total $\kappa$, mainly because the measurements have not been carried out at low enough temperatures.\cite{Sun_DTN, Kohama} In the case of Ba$_3$Mn$_2$O$_8$, the magnons in the BEC state are scattering phonons rather than transporting heat.\cite{Ke_BMO} In this work, we choose another BEC candidate (CH$_3$)$_2$CHNH$_3$CuCl$_3$ (IPA-CuCl$_3$)\cite{IPA-1, IPA-2, IPA-3, IPA-4, IPA-5, IPA-6, Manaka} to study the thermal conductivity. IPA-CuCl$_3$ crystallizes in a triclinic structure and the Cu$^{2+}$ spins ($S$ = 1/2) form ladders along the $a$ axis, with rungs along the $c$ axis, as shown in Fig. 1. The zero-field ground state is quantum paramagnetic with a spin gap of 1.17 meV. When the magnetic field closes the gap at $\mu_0H_{c1} \sim$ 10 T, an AF state is developed. This has been proposed as a BEC state since the neutron scattering indicated a gapless mode.\cite{IPA-3, IPA-4} Here, we study the heat transport of IPA-CuCl$_3$ single crystal at very low temperatures down to several tens of millikelvin. At $H = H_{c1}$, a $T^3$ magnon thermal conductivity is observed at $T <$ 600 mK, which is the clearest experimental evidence of the ballistic magnon heat transport in the AF state till now. However, the low-$T$ $\kappa(H)$ isotherms indicate that the spin-gap is closed only at $H_{c1}$. The re-opening of a small gap at $H > H_{c1}$ demonstrates that the BEC model is not strictly applicable to IPA-CuCl$_3$.

\begin{figure}
\includegraphics[clip,width=7.5cm]{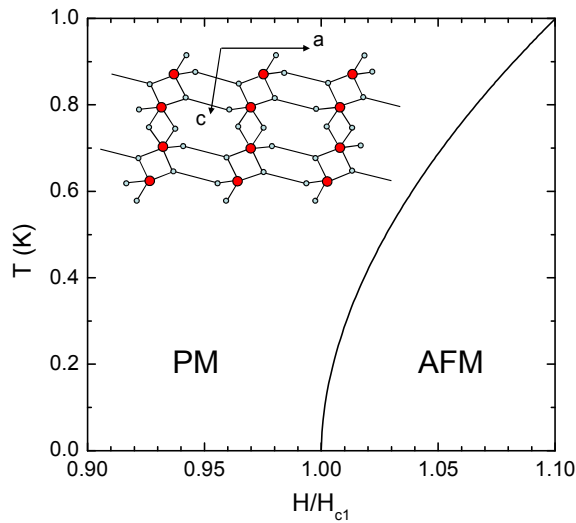}
\caption{(Color online) Schematic $H-T$ phase diagram of IPA-CuCl$_3$ according to the magnetization, specific heat, magnetocaloric effect, and neutron scattering measurements.\cite{IPA-1, IPA-3, IPA-4, IPA-5, Manaka} PM represents the low-field quantum paramagnetic state and AFM represents the field-induced antiferromagnetically ordered state, which has been discussed to be a possible BEC state. The solid line is the phase boundary ($H_{c1}$) of the above two states. Note that the critical fields ($\sim$ 10 T at $T \rightarrow$ 0) are slightly different for different field directions and from different experiments. Inset: Schematic plot of the $ac$-plane structure, including the magnetic Cu$^{2+}$ (big and red circles) and the bridging Cl$^-$ (small and cyan circles) ions. Other atoms are omitted for clarity.}
\end{figure}

\section{EXPERIMENTS}

IPA-CuCl$_3$ single crystals were grown by a slow evaporation of ethanol solution mixed with CuCl$_2$, isopropyl amine, and concentrated HCl in an appropriate proportion.\cite{JACS} The as-grown crystals are dark brown with size up to $15 \times 5 \times 0.6$ mm$^3$, with the length and width nearly along the $c$ and $a$ axis, respectively. The specific heat of an IPA-CuCl$_3$ single crystal was measured by the relaxation method in the temperature range from 2 to 30 K using a commercial physical property measurement system (PPMS, Quantum Design). For anisotropic $\kappa$ measurements, the long-bar shaped samples were cut from the as-grown crystals along either the $a$ or $c$ axis, respectively. The $\kappa$ was measured using a ``one heater, two thermometers" technique in a $^3$He refrigerator at 300 mK $< T <$ 30 K and a $^3$He-$^4$He dilution refrigerator at 70 mK $< T <$ 1 K, equipped with a 14 T magnet.\cite{Zhao_DFO, Zhou_ZCO, Sun_YBCO} In these measurements, the magnetic field is parallel to the heat current ($J \rm_H$), which is along the lengths of the samples. It should be noted that the crystals are so fragile that cutting and polishing can easily produce some damages inside the samples; furthermore, some samples were even broken after cooling and warming cycle. Thus, the data presented in this paper were taken on several different samples.

\section{RESULTS}

\begin{figure}
\includegraphics[clip,width=6.5cm]{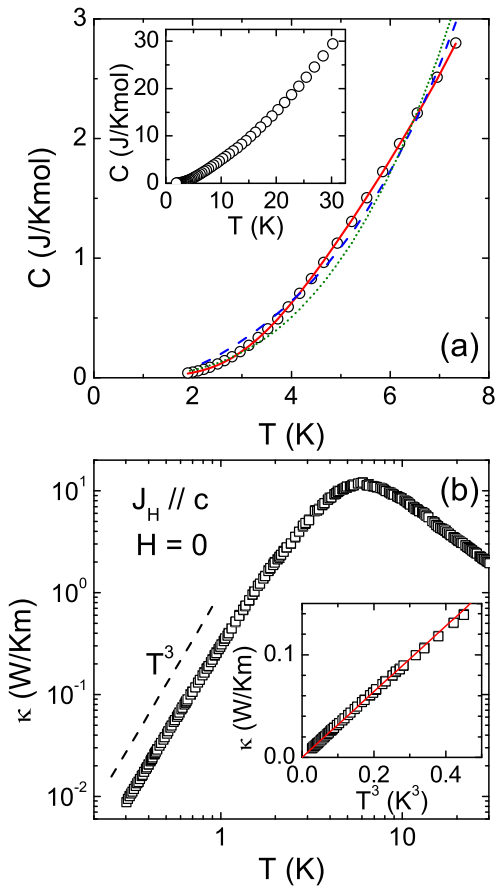}
\caption{(Color online) (a) Low-temperature specific heat of an IPA-CuCl$_3$ single crystal in zero field. The solid line is a fitting to the data using Eq. (\ref{SH}) with adjustable parameters $\beta =$ 4.81 $\times$ 10$^{-3}$ J/K$^4$mol, $\Delta/k_B$ = 22.1 K and $A =$ 0.88 J/Kmol. The dashed line is a different fitting using Eq. (\ref{SH}) with the fixed parameter $\Delta/k_B$ = 13.6 K and adjustable parameters $\beta =$ 7.18 $\times$ 10$^{-3}$ J/K$^4$mol and $A =$ 0.19 J/Kmol. The dotted line is a simple $T^3$ fitting with coefficient of 7.92 $\times$ 10$^{-3}$ J/K$^4$mol. The inset shows data in a broader temperature range from 2 to 30 K. (b) Temperature dependence of thermal conductivity of an IPA-CuCl$_3$ single crystal in zero field and at 300 mK -- 30 K. The heat current is applied along the $c$ axis. The dimension of this sample is $4.6 \times 1.74 \times 0.88$ mm$^3$. The dashed line indicates $T^3$ temperature dependence. Inset: the low-temperature data in a linear plot for $\kappa$ vs $T^3$. The thin line is a fitting to $\kappa = bT^3$ with the parameter $b$ = 0.32 W/K$^4$m for $T <$ 700 mK.}
\end{figure}

Figure 2(a) shows the specific heat of an IPA-CuCl$_3$ single crystal with 2 $\le T \le$ 30 K. Our data are consistent with the reported data.\cite{Manaka} It is notable that the temperature dependence strongly deviates from the well-known $T^3$-behavior of phonon specific heat. A fitting of low-$T$ data to $C_p = \beta T^3$ is shown by a dotted line in Fig. 2(a). Furthermore, the data cannot be well fitted by a more complicated formula of phononic specific heat,
\begin{equation}
C_p = \beta T^3 + \beta_5 T^5+\beta_7 T^7,
\label{PSH}
\end{equation}
which is the low-frequency expansion of the Debye function with $\beta$, $\beta_5$ and $\beta_7$ the $T$-independent coefficients.\cite{Tari} The reason is mainly due to the contribution of magnetic excitations. As suggested by Ref. \onlinecite{Manaka}, since the existence of a spin gap of magnetic excitations, the low-$T$ specific heat can be described as a $T^3$ phononic term plus a Schottky term, that is,
\begin{equation}
C = \beta T^3 + \frac{1}{3} A \left(\frac{\Delta}{k_BT}\right)^{2} \frac{e^{\Delta/k_BT}}{(1+(1/3)e ^{\Delta/k_BT})^{2}},
\label{SH}
\end{equation}
where $\Delta$ is the spin gap separating the singlet ground state and the triplet first-excitation state, $A$ is an adjusting parameter. The factor 1/3 is related to the triple degeneracy of the first-excitation state.\cite{Tari} This formula can fit the low-$T$ specific heat data well with the parameters $\beta =$ 4.81 $\times$ 10$^{-3}$ J/K$^4$mol, $\Delta/k_B$ = 22.1 K and $A =$ 0.88 J/Kmol, as shown in Fig. 2(a). These parameters are also consistent with the results in the earlier literature.\cite{Manaka} However, the fitting parameter $\Delta$ is rather different from the size of the spin gap (13.6 K) obtained from neutron measurements.\cite{IPA-2} If we fit the data using formula (\ref{SH}) with a fixed parameter $\Delta/k_B$ = 13.6 K, the fitting is much worse, as shown by the dashed line in Fig. 2(a). This discrepancy should be due to the complexity of low-$T$ specific heat. Some other factors, such as the spin disorders or magnetic/nonmagnetic impurities, may give small additional contributions to the specific heat.

Figure 2(b) shows the temperature dependence of $\kappa$ of an IPA-CuCl$_3$ single crystal in zero field with the heat current along the $c$ axis. Due to the spin gap of 1.17 meV (13.6 K) in the ground state of IPA-CuCl$_3$,\cite{IPA-1, IPA-2} the magnetic excitations can hardly be thermally excited at very low temperatures. Therefore, the low-$T$ $\kappa(T)$ is the pure phonon conductivity. Indeed, the low-$T$ $\kappa(T)$ behaves as a common insulator, with a phonon peak locating at about 6 K. The magnitude of peak exceeds 10 W/Km and is comparable with some other organic low-dimensional magnets.\cite{Sun_DTN, Chen_MCCL, Sologubenko1}

A notable phenomenon is that the $\kappa(T)$ follows a perfect $T^3$ dependence at $T <$ 700 mK. In principle, the phonon thermal conductivity can be expressed as a kinetic formula $\kappa_{p} = \frac{1}{3}C v_p l_p$, in which $C = \beta T^3$ is the low-$T$ specific heat, $v_p$ is the averaged sound velocity and is nearly $T$-independent at low temperatures, and $l_p$ is the mean free path of phonons.\cite{Berman} With decreasing temperature, the microscopic scattering of phonons are gradually smeared out and the $l_p$ increases continuously untill reaching the averaged sample width $W = 2\sqrt{A/\pi}$, where $A$ is the cross-section area of sample.\cite{Berman} This boundary scattering limit of phonons can be achieved only at very low temperatures and the $T$-dependence of $\kappa_p$ is the same as the $T^3$ law of the specific heat.\cite{Berman, Ziman} The well-known $T^3$ ballistic behavior of phonons, however, has been rarely observed in the transition-metal compounds, including the high-$T_c$ cuprates, the multiferroic manganites and the low-dimensional quantum magnets.\cite{Taillefer, Sun_YBCO, Sun_Nonuniversal, Sologubenko1, Sologubenko2, Yamashita, Zhao_DFO, Sun_DTN, Ke_BMO, Chen_MCCL, Zhou_ZCO}

In the boundary scattering limit, the phonon thermal conductivity of an isotropic system is given by\cite{Ziman}
\begin{equation}
\kappa_{p} = \frac{2\pi^2}{15} \frac{l_{p}}{v_p^2}k_B \left(\frac{k_B}{\hbar}\right)^3 T^3. \label{kp}
\end{equation}
The averaged phonon velocity $v_p$ can be extracted from the specific-heat coefficient using the relations $\beta = \frac{12\pi^4}{5} \frac{Rs}{\Theta_D^3}$ and $\Theta_D = \frac{\hbar v_p}{k_B} (\frac{6\pi^2 Ns}{V})^\frac{1}{3}$,\cite{Tari} where $\Theta_D$ is the Deybe temperature, $N$ is the number of molecules per mole and each molecule comprises $s$ atoms, $V$ is the volume of crystal and $R$ is the universal gas constant. For phonons in a three-dimensional (3D) lattice, the anisotropy of phonon velocity is usually not very strong and formula (\ref{kp}) can describe the $\kappa_p$ rather well.

The present low-$T$ data is fitted well to $\kappa = bT^3$ with $b$ = 0.32 W/K$^4$m, as shown in the inset to Fig. 2(b). Thus, with the parameter $\beta =$ 4.81 $\times$ 10$^{-3}$ J/K$^4$mol, the mean free path is calculated to be 0.017 mm. This value is much smaller than the geometry size of this sample, $W =$ 1.40 mm. Since the $T^3$ behavior is a robust signature of the boundary scattering limit, it seems that the sample has some small cracks that act as the boundaries of phonon transport. As mentioned above, it is likely that this kind of cracks are not intrinsic but are produced in the cutting, polishing and cooling processes.

\begin{figure}
\includegraphics[clip,width=6.5cm]{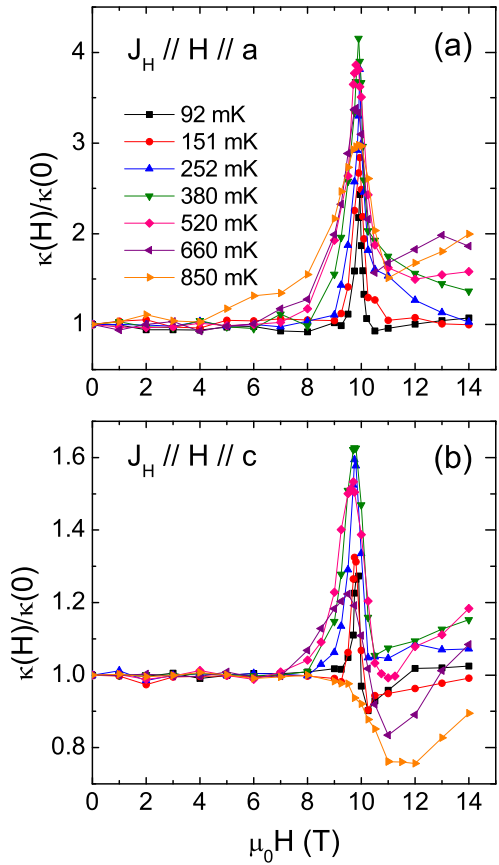}
\caption{(Color online) Magnetic-field dependencies of the $\kappa_a$ (a) and $\kappa_c$ (b) of two IPA-CuCl$_3$ crystals at subkelvin temperatures. Since the magnetic fields are applied along the direction of the heat current, the demagnetization effect is negligibly small for these long-bar shaped samples.}
\end{figure}

Figure 3 shows the magnetic-field dependencies of $\kappa_a$ and $\kappa_c$ at low temperatures for two IPA-CuCl$_3$ single crystals. The data exhibit two remarkable features. First, the $\kappa$ are field independent at low fields but exhibit a sharp peak at $\mu_0H_{c1} \sim $ 9.95 and 9.75 T for $H \parallel a$ and $H \parallel c$, respectively, particularly at very low temperatures. It is noted that at $T \rightarrow$ 0 the peak positions are coincided with the transition field of the field-induced AF state.\cite{IPA-1, IPA-3, IPA-4, IPA-5, Manaka} Since the magnons can be easily excited when the spin gap is closed at $H_{c1}$, the increase of the $\kappa$ at $H_{c1}$ is a direct evidence of the magnon heat transport. Note that the small anisotropy (9.95/9.75 = 1.02) of critical fields between $H \parallel a$ and $H \parallel c$ is due to the anisotropy of $g$ values. The electron spin resonance (ESR) measurements showed that the $g$ values along the $a$ and $c$ axes are 2.05 and 2.11, respectively.\cite{Manaka2} This anisotropy (2.11/2.05 = 1.03) explains the anisotropy of $H_{c1}$. Second, the peak feature demonstrates that the gap is closed only at $H_{c1}$ and is re-opened above $H_{c1}$, which results in the vanishing of magnon transport in high fields.

\begin{figure}
\includegraphics[clip,width=8.5cm]{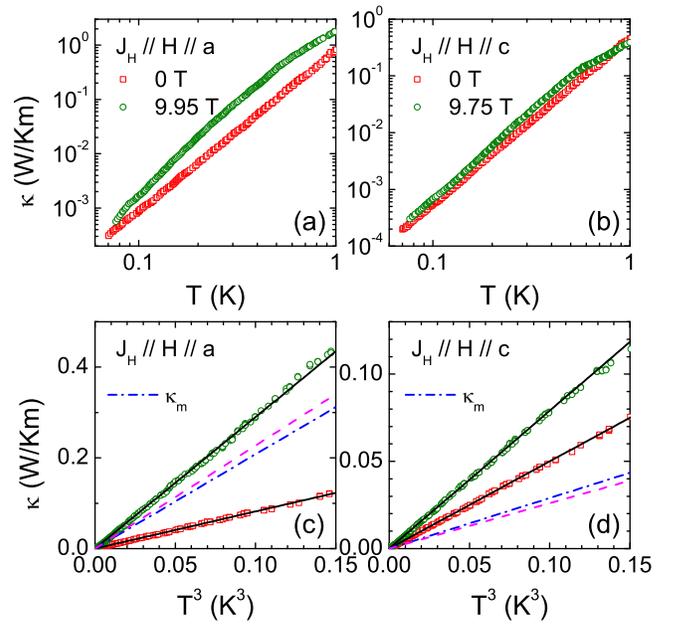}
\caption{(Color online) (a) Temperature dependencies of the $\kappa_a$ in zero field and 9.95 T ($\parallel a$). (b) Temperature dependencies of the $\kappa_c$ in zero field and 9.75 T ($\parallel c$). The data are taken on two IPA-CuCl$_3$ crystals, of which the $\kappa(H)$ data are shown in Fig. 3. The dimensions of the $\kappa_a$ and $\kappa_c$ samples are $4.5 \times 1.29 \times 0.55$ mm$^3$ and $3.91 \times 1.54 \times 0.55$ mm$^3$, respectively. (c,d) Data shown with $\kappa$ vs $T^3$ plot. The solid lines are linear fittings to the experimental data. The dot dashed lines denote the magnon thermal conductivity, $\kappa_{m,a} = 2.08T^3$ (W/Km) and $\kappa_{m,c} = 0.29T^3$ (W/Km), obtained by subtracting the zero-field data from the $\kappa(T)$ curves in the critical fields. The dashed lines show the calculated curves $\kappa_{m,a} = 2.27T^3$ (W/Km) and $\kappa_{m,c} = 0.26T^3$ (W/Km), using formula (\ref{kmi}) with appropriate parameters, that are closest to the experimental data of $\kappa_{m,a}$ and $\kappa_{m,c}$.}
\end{figure}

To probe the $T$-dependence of the magnon heat transport, the $\kappa_a(T)$ and $\kappa_c(T)$ at $H_{c1}$ of these two samples are measured and compared with their zero-field data, as shown in Fig. 4. Both the $\kappa_a$ and $\kappa_c$ in $H =$ 0 follow the $\kappa = bT^3$ dependence at low temperatures, with $b =$ 0.82 W/K$^4$m for $\kappa_a$ and 0.50 W/K$^4$m for $\kappa_c$, respectively. The phonon mean free paths in the ballistic regime are calculated to be 0.043 and 0.026 mm for the $\kappa_a$ and $\kappa_c$ samples, respectively, which are also smaller than the sample sizes ($W =$ 0.94 and 1.03 mm). All these are consistent with the result of another sample shown in Fig. 2.

The most important finding is that the $\kappa_a$ ($\kappa_c$) at 9.95 (9.75) T also follow the $T^3$ dependence at $T <$ 600 (700) mK. The $\kappa = aT^3$ fittings yield $a =$ 2.90 W/K$^4$m for $\kappa_a$ in 9.95 T and 0.79 W/K$^4$m for $\kappa_c$ in 9.75 T, respectively. Therefore, the magnon thermal conductivity at $H_{c1}$ can be obtained by subtracting the zero-field data from the critical-field curves, which gives $\kappa_m = 2.08 T^3$ (W/Km) and $0.29T^3$ (W/Km) along the $a$ and $c$ axis, respectively, as shown in Figs. 4(c) and 4(d). This nearly perfect $T^3$ dependence is actually the clearest experimental evidence of the magnon ballistic transport in the AF systems till now.

There are several notable details about the field- and temperature-dependencies of $\kappa$. First, for both field directions, the transition field determined by the peak positions of $\kappa(H)$ isotherms is nearly temperature independent and slightly shifts to lower fields with increasing temperature. This $T$-dependence of $H_{c1}$ is actually not the same as the transition field closing the spin gap probed by other measurements.\cite{IPA-1, IPA-3, IPA-4, IPA-5, Manaka} Very similar phenomenon has been found in another BEC material DTN.\cite{Sun_DTN} The possible reason is that the magnetic excitations may not only transport heat but also scatter phonons. Since these two factors contribute oppositely to $\kappa$, the field dependence of $\kappa$ is rather complicated and the maximum could appear at the fields slightly different from the phase-transition fields. In particular, this discrepancy is stronger at higher temperatures where the scattering between phonons and magnons are stronger and the peak becomes broader. Second, at the critical fields, $\kappa(T)$ show some deviations of the $T$-dependence from the exact $T^3$ behavior at both high and low temperatures. At high temperature ($>$ 500 mK), the $T$-dependence becomes a bit weaker, which is a usual phenomenon of heat transport if the scattering between magnetic excitations and phonons cannot be neglected.\cite{Wang_HMO, Zhao_GFO, Zhao_DFO, Sun_DTN, Ke_BMO} That is, it is a simple deviation from the boundary scattering limit. Another possible reason for this deviation is a dimensional crossover of spin system to the two-dimensional (2D) regime, as found, for example, in a 2D antiferromagnet RbFe(MoO$_4$)$_2$.\cite{Svistov} In this regard, there seemed to be no experimental result indicating such a dimensional crossover in IPA-CuCl$_3$.\cite{IPA-1, IPA-3, IPA-4, IPA-5, Manaka} On the other hand, $\kappa_a(T)$ at 9.95 T shows stronger $T$-dependence than $T^3$ at the lowest temperature regime. So far, the reason is not very clear. Probably some decoupling between phonons and magnons occurs, like the electron-phonon decoupling in high-$T_c$ cuprates.\cite{Smith, Zhao_NCCO}

\section{Data analysis and discussions}

Similar to the case of phonons, the ballistic magnon thermal conductivity of an isotropic system can be written as \cite{Berman, Ziman, Ashcroft}
\begin{equation}
\kappa_m = \frac{1}{3} \times \frac{2}{15} \pi^2k_B \left(\frac{k_BT}{\hbar}\right)^3 v_m^{-2} l_m, \label{km}
\end{equation}
where $v_m$ is the averaged magnon velocity, and $l_m$ is the $T$-independent mean free path of magnons (see Appendix). It is easy to find that Eq. (\ref{km}) of the isotropic system cannot quantitatively describe the experimental results of $\kappa_{m,a}$ and $\kappa_{m,c}.$ Note that at ultra-low temperatures, the $l_m$ is also determined by the boundary scattering and is actually the same as the mean free path of phonons. The calculation using Eq. (\ref{km}) gives the magnon velocities of 920 and 1930 m/s along the $a$ and the $c$ axis, respectively. They are apparently unreasonable since the magnon dispersion is stronger along the $a$ axis (the ladder direction).\cite{IPA-1, IPA-2, IPA-3, IPA-4, IPA-5, IPA-6} It is simply due to the fact that Eq. (\ref{km}) is established for the isotropic system and cannot be valid for the low-dimensional spin systems. As shown in the Appendix, for the anisotropic system the magnon thermal conductivity along a certain direction $i$ is expressed as
\begin{equation}
\kappa_{m,i} = \frac{1}{3} \times \frac{\pi^3}{30} k_B \left(\frac{k_BT}{\hbar}\right)^3 \bar{v}_{m,i} l_{m,i} \alpha' , \label{kmi}
\end{equation}
where $\bar{v}_{m,i}$, $l_{m,i}$ and $\alpha'$ represent the averaged magnon velocity along the direction $i$, the $T$-independent mean free path and the coefficient of frequency density distribution, respectively. Here, $\alpha'$ is written as
\newcommand{\ud}{\mathrm{d}}
\begin{equation}
\alpha' = \frac{\int_0^{\pi/2}\ud\theta \int_0^{\pi/2}W(\theta,\phi)v^{-3}(\theta,\phi)\sin\phi\ud\phi}
{\int_0^{\pi/2}\ud\theta\int_0^{\pi/2}W(\theta,\phi)\ud\phi}, \label{beta}
\end{equation}
while $W(\theta,\phi)$ is an assumed velocity-weighting function with an exponential form,
\begin{equation}
W(\theta,\phi) = e^{\frac{v(\theta,\phi)}{v_0}}, \label{W}
\end{equation}
with $v_0$ an adjustable parameter, $\theta$ and $\phi$ the space angles.

The magnon dispersions along the $a$ and $c$ axis of IPA-CuCl$_3$ have been known from the earlier neutron measurements, which gave $v_{m,a} =$ 2050 m/s and $v_{m,c}$ = 790 m/s.\cite{IPA-2, IPA-3, IPA-4} The magnon dispersion is known to be very weak along the $b$ axis.\cite{IPA-2} Here, we use Eq. (\ref{kmi}) to calculate the $\kappa_{m,a}$ and $\kappa_{m,c}$ with the experimental values of $v_{m,a}$ and $v_{m,c}$ and adjustable parameters $v_{m,b}$ and $v_0$. According to this formula, these two parameters have different impacts on the value of $\kappa_{m,a}$ and $\kappa_{m,c}$. The magnitudes of $\kappa_{m,a}$ and $\kappa_{m,c}$ increase with decreasing $v_{m,b}$ and their ratio (or anisotropy) hardly changes with $v_{m,b}$. In contrast, $\kappa_{m,a}$ and $\kappa_{m,c}$ decrease with decreasing $v_0$. Furthermore, the ratio of $\kappa_{m,a} / \kappa_{m,c}$ increases with decreasing $v_0$. Therefore, only with appropriate values of $v_{m,b}$ and $v_0$, one can get calculated $\kappa$ close to the experimental data for both $\kappa_{m,a}$ and $\kappa_{m,c}$ simultaneously. The best calculations of $\kappa_{m,a} = 2.27T^3$ (W/Km) and $\kappa_{m,c} = 0.26T^3$ (W/Km), as shown in Figs. 4(c) and 4(d), are obtained with $v_{m,b} =$ 150 m/s and $v_0 =$ 220 m/s. Note that the parameter $v_{m,b}$ is about one order of magnitude smaller than the velocities along the other axes, which is reasonable for IPA-CuCl$_3$.\cite{IPA-2, IPA-3, IPA-4} Therefore, the ballistic magnon heat transport of a low-dimensional quantum magnet can be well described by an anisotropic formula (\ref{kmi}).

Another important result of this work is the peak-like feature of $\kappa(H)$ at $H_{c1}$, which becomes clearer and sharper with lowering temperature. As already mentioned above, this indicates that the spin gap is closed only at the critical field and is re-opened in the field-induced AF state. Therefore, the lowest excitation in the field-induced AF state is non-Goldstone mode. The gap size cannot be determined precisely by the present data, but we can take a rough estimation from the high-field behavior of $\kappa(H)$. In the high-field state with small gap, the magnons can still be easily excited and contribute to transporting heat if $k_BT$ is not smaller than the gap. As shown in Fig. 3, it seems that the $\kappa$ tends to recover its zero-field value at high-field limit of $H \parallel a$ ($H \parallel c$ ) when $T \le$ 380 mK (252 mK) but tends to increase at high-field limit when $T \ge$ 520 mK (380 mK). Therefore, the gap is estimated to be about 500 mK ($\approx$ 0.043 meV) and 300 mK ($\approx$ 0.026 meV) for $H \parallel a$ and $c$, respectively. Apparently, such small gaps are beyond the resolution of the earlier neutron measurements. Similarly, TlCuCl$_3$ was the first BEC candidate that showed a gapless Goldstone mode in the field-induced AF state by the neutron scattering measurement,\cite{BEC-1} but was lately found to have a small gap of 0.09 meV by the ESR measurement.\cite{TlCuCl3-1, TlCuCl3-2}

A characteristic of BEC is the presence of $U(1)$ symmetry, which corresponds to the global rotational symmetry of the bosonic field phase.\cite{Review-1, Review-2} In field-induced XY-ordered state, the $U(1)$ symmetry spontaneously gets broken and thus a gapless Goldstone mode is acquired. However, the re-opening of the gap is a clear evidence for a broken uniaxial symmetry of spin Hamiltonian, which rules out a strict description of the magnetic order in terms of BEC.\cite{Review-2} Therefore, the heat transport data indicate that the BEC model has limited applicability to IPA-CuCl$_3$, similar to TlCuCl$_3$. The theoretical works actually had predicted a general instability of an axially symmetric magnetic condensate toward a violation of this symmetry and the formation of an anisotropy gap at $H > H_{c1}$.\cite{Review-2, Amore} It is related to the presence of anisotropic interactions, such as the dipole-dipole coupling, the spin-orbital interaction, etc.\cite{Review-2, Amore}

\section{Conclusions}

In summary, we have studied the ultra-low-$T$ heat transport of a spin-gapped compound IPA-CuCl$_3$, which has been classified to be a candidate of BEC. When the gap is closed by the field at $H_{c1}$, a $T^3$ ballistic magnon heat transport is observed. On the other hand, the low-$T$ $\kappa(H)$ isotherms show peaks at the critical field, indicating that the spin gap is re-opened at $H > H_{c1}$. Therefore, IPA-CuCl$_3$ seems not to be an ideal BEC prototype system. Ultra-low-$T$ thermal conductivity is a very sensitive technique to probe the small spin gap, and the validity of BEC model for those quantum magnets can be carefully examined using this measurement.

\begin{acknowledgements}

This work was supported by the National Natural Science Foundation of China, the National Basic Research Program of China (Grant Nos. 2015CB921201 and 2011CBA00111), and the Fundamental Research Funds for the Central Universities (Program No. WK2030220014).

\end{acknowledgements}

\setcounter{equation}{0}
\renewcommand{\theequation}{A\arabic{equation}}

\section*{Appendix: Phenomenological formulas for the heat transport of phonons and magnons}

\subsection{The isotropic systems}

According to the Boltzmann equation, the lattice thermal conductivity can be obtained by solving the integral equation. In an isotropic (or cubic) 3D crystal, the phononic thermal conductivity is given by\cite{Berman, Ziman, Ashcroft, Holland}
\begin{equation}\label{eq:eps}
\kappa_{p} = \frac{1}{3}\frac{1}{(2\pi)^3} \sum_\lambda \int v_{p,\lambda}(q) l_{p,\lambda}(q) C_{p,\lambda}(q) f_{\lambda}(q) \mathrm{d}q,
\end{equation}
where $v_{p,\lambda}(q)$, $l_{p,\lambda}(q)$, $C_{p,\lambda}$ and $f_{\lambda}(q)$ are the velocity, mean free path, specific heat per normal mode and vibration mode distribution function of phonons with polarization $\lambda$ (two transverse and one longitudinal) and wave vector $q$, respectively. The specific heat $C_{p,\lambda}$ is expressed as
\begin{equation}\label{eq:eps}
C_{p,\lambda}(q) = \frac{\mathrm{d}}{\mathrm{d}T} \left(\hbar\omega_{\lambda}(q)\cdot\frac{1}{e^{\frac{\hbar\omega_{\lambda}(q)}{k_BT}}-1}\right),
\end{equation}
where $\omega_{\lambda}(q)$ is the dispersion for polarization $\lambda$.

Usually, the following assumptions are needed to simplify the equation and calculate the result. First, it is assumed that phonons of different polarizations have the same contribution, which means $v_{p}(q)$, $l_{p}(q)$, $C_{p}$, $f(q)$ and $\omega(q)$ are independent of polarization $\lambda$. Second, according to the isotropic Debye approximation, $v_p$ is a constant and the dispersion relation is $q = \omega/v_p$ in all directions. Thus, $q$ is independent of direction and $f(q)\mathrm{d}q = 4\pi q^2\mathrm{d}q = 4\pi \omega^2v_p^{-3}\mathrm{d}\omega$. With these assumptions, the thermal conductivity can be simplified as
\begin{equation}\label{eq:eps}
\kappa_{p} =\frac{1}{2\pi^2} \frac{l_{p}}{v_p^2}k_B \left(\frac{k_B}{\hbar}\right)^3 T^3\int_0^{\Theta_D /T} \frac{x^4e^x}{(e^x-1)^2}\mathrm{d}x,
\end{equation}
where $x = \frac{\hbar\omega}{k_BT}$. If the temperature is low enough, there is only boundary scattering and the mean free path is a constant given by the cross-section area, that is, $l_{p} = 2\sqrt{A/\pi}$. In this case, the integral approaches a constant of $4\pi^4/15$ and thermal conductivity obeys a $T^3$ law, which is formula (\ref{kp}) mentioned in the main text.

This result can also be applied to gapless acoustic magnons of a 3D antiferromagnet. Note that when the degeneracy of gapped magnon branches is lifted by magnetic field, only the lowest branch becomes gapless. Hence, the magnon thermal conductivity is
\begin{equation}\label{eq:eps}
\kappa_{m} = \frac{1}{3} \times \frac{2\pi^2}{15} \frac{l_{m}}{v_m^2}k_B \left(\frac{k_B}{\hbar}\right)^3 T^3,
\end{equation}
where $v_m$ and $l_m$ are the velocity and mean free path of magnons.

\subsection{The anisotropic systems}

The above formulas for phonons and magnons are valid only for the isotropic systems. Usually, the anisotropy of phonons is not strong for most 3D crystal lattices and therefore formula (\ref{kp}) can describe the phonon heat transport in a good approximation. For anisotropic systems, these formulas predict that the $\kappa$ is smaller along the direction with stronger dispersion (larger velocity). This is apparently unreasonable. For strongly anisotropic systems, like the magnons in a low-dimensional quantum magnet, this problem would be more serious. Here, we give some phenomenological results of the Boltzmann theory in the anisotropic systems.

\begin{figure}
\includegraphics[clip,width=6.5cm]{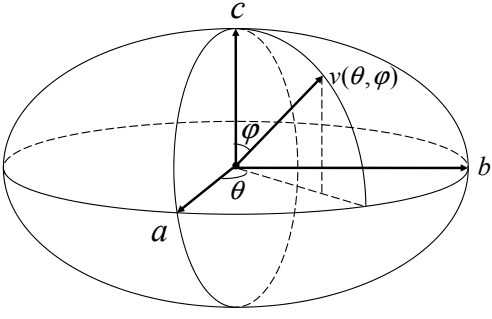}
\caption{Schematic of the ellipsoid velocity distribution. }
\end{figure}

Let us again start with the phonon heat transport. Note that with the above assumptions for isotropic system,\cite{Berman, Ziman, Ashcroft, Holland} phonons only travel along the temperature gradient direction with a uniform velocity. In the isotropic case, we need to consider phonons traveling in all directions with component along the temperature gradient. Here, the phonon velocity is a function of space angles and a simple function model is an ellipsoid (see Fig. 5). Thus, the projections of a random velocity of one acoustic branch can be written in parametric form as
\begin{equation}\label{eq:eps}
v_a = a\cos\sigma\cos\gamma,
\end{equation}
\begin{equation}\label{eq:eps}
v_b = b\sin\sigma\cos\gamma,
\end{equation}
\begin{equation}\label{eq:eps}
v_c = c\sin\gamma.
\end{equation}
Here, $a$, $b$ and $c$ are the phonon velocities along the three principal crystal axes. $\sigma$ and $\gamma$ are parameters of the ellipsoid equation and they are represented by space angles $\theta$ and $\phi$ as
\begin{equation}\label{eq:eps}
\sigma = \arctan \left( \frac{a\tan\theta}{b} \right),
\end{equation}
\begin{equation}\label{eq:eps}
\gamma = \arctan\left( \frac{\sqrt{(a\cos\sigma)^2+(b\sin\sigma)^2}}{c \tan\phi}\right).
\end{equation}
Then, the anisotropic sound velocity can be written as a function of space angles,
\begin{equation}\label{eq:eps}
v(\theta,\phi) = \sqrt{{v_a}^2(\theta,\phi)+{v_b}^2(\theta,\phi)+{v_c}^2(\theta,\phi)}.
\end{equation}

In this anisotropic velocity model, $v_{p,q}$ and $f(q)\mathrm{d}q$ need correction. With Eq. (A10), the averaged phonon velocity along the $a$, $b$ and $c$ axis are written as
\begin{equation}\label{eq:eps}
\bar{v}_a = \frac{\int_0^{\pi/2}\mathrm{d}\theta\int_0^{\pi/2} v(\theta,\phi)\cos\theta\sin\phi\mathrm{d}\phi}{\int_0^{\pi/2}\mathrm{d}\theta\int_0^{\pi/2}\mathrm{d}\phi},
\end{equation}
\begin{equation}\label{eq:eps}
\bar{v}_b = \frac{\int_0^{\pi/2}\mathrm{d}\theta\int_0^{\pi/2} v(\theta,\phi)\sin\theta\sin\phi\mathrm{d}\phi}{\int_0^{\pi/2}\mathrm{d}\theta\int_0^{\pi/2}\mathrm{d}\phi},
\end{equation}
\begin{equation}\label{eq:eps}
\bar{v}_c = \frac{\int_0^{\pi/2}\mathrm{d}\theta\int_0^{\pi/2} v(\theta,\phi)\cos\phi\mathrm{d}\phi} {(\pi/2) \int_0^{\pi/2}\mathrm{d}\theta\int_0^{\pi/2}\mathrm{d}\phi}.
\end{equation}

Since $v(\theta,\phi)$ depends on the space angles here, $q = \omega/v(\theta,\phi)$ is function of $\omega$, $\theta$ and $\phi$ simultaneously. Thus, $f(q)\mathrm{d}q$ has a more complex form as
\begin{equation}\label{eq:eps}
f(q)\mathrm{d}q = 4\alpha\omega^2\mathrm{d}\omega,
\end{equation}
where
\begin{equation}\label{eq:eps}
\alpha = \int_0^{\pi/2}\mathrm{d}\theta\int_0^{\pi/2} v^{-3}(\theta,\phi) \sin\phi\mathrm{d}\phi.
\end{equation}
Take Eqs. (A2) and (A11-A14) into Eq. (A1), one can get the thermal conductivities along direction $i$ for phonons and magnons in low temperature limit as
\begin{equation}\label{eq:eps}
\kappa_{p,i} = \frac{2\pi}{15}\bar{v}_{p,i}l_{p,i}\alpha k_B\left(\frac{k_B}{\hbar}\right)^3 T^3
\end{equation}
and
\begin{equation}\label{eq:eps}
\kappa_{m,i} = \frac{1}{3}\times\frac{2\pi}{15}\bar{v}_{m,i}l_{m,i}\alpha k_B\left(\frac{k_B}{\hbar}\right)^3 T^3,
\end{equation}
respectively.

Here, we put forward a further quantitative assumption involving the anisotropy. We consider that phonons or magnons prefer to propagate in a direction with stronger dispersion and this direction of course contributes more to the average velocity. Thus, we assume a velocity-dependent weighting function with an exponential form,
\begin{equation}\label{eq:eps}
W(\theta,\phi) = e^{\frac{v(\theta,\phi)}{v_0}},
\end{equation}
where $v_0$ is an adjustable parameter. With this weighting function, the averaged phonon velocities in direction $a$, $b$ and $c$ should be written as
\begin{equation}\label{eq:eps}
\bar{v}_a = \frac{\int_0^{\pi/2}\mathrm{d}\theta\int_0^{\pi/2}W(\theta,\phi)v(\theta,\phi)\cos\theta\sin\phi\mathrm{d}\phi} {\int_0^{\pi/2}\mathrm{d}\theta\int_0^{\pi/2}W(\theta,\phi)\mathrm{d}\phi},
\end{equation}
\begin{equation}\label{eq:eps}
\bar{v}_b = \frac{\int_0^{\pi/2}\mathrm{d}\theta\int_0^{\pi/2}W(\theta,\phi)v(\theta,\phi)\sin\theta\sin\phi\mathrm{d}\phi} {\int_0^{\pi/2}\mathrm{d}\theta\int_0^{\pi/2}W(\theta,\phi)\mathrm{d}\phi},
\end{equation}
\begin{equation}\label{eq:eps}
\bar{v}_c = \frac{\int_0^{\pi/2}\mathrm{d}\theta\int_0^{\pi/2}W(\theta,\phi)v(\theta,\phi)\cos\phi\mathrm{d}\phi} {(\pi/2)\int_0^{\pi/2}\mathrm{d}\theta\int_0^{\pi/2}W(\theta,\phi)\mathrm{d}\phi}.
\end{equation}
The denominators are normalization coefficients.

For the same reason, taking the weighting function into account, $f(q)\mathrm{d}q$ has a similar form of
\begin{equation}\label{eq:eps}
f(q)\mathrm{d}q = \pi^2\alpha'\omega^2\mathrm{d}\omega,
\end{equation}
and here
\begin{equation}\label{eq:eps}
\alpha' = \frac{\int_0^{\pi/2}\mathrm{d}\theta \int_0^{\pi/2}W(\theta,\phi)v^{-3}(\theta,\phi)\sin\phi\mathrm{d}\phi}
{\int_0^{\pi/2}\mathrm{d}\theta\int_0^{\pi/2}W(\theta,\phi)\mathrm{d}\phi}.
\end{equation}
Take Eqs. (A2) and (A19-A22) into Eq. (A1), one can finally obtain the phonon and magnon thermal conductivities along direction $i$ ($a$, $b$ or $c$) as
\begin{equation}\label{eq:eps}
\kappa_{p,i} = \frac{\pi^3}{30}\bar{v}_{p,i}l_{p,i}\alpha' k_B\left(\frac{k_B}{\hbar}\right)^3 T^3
\end{equation}
and
\begin{equation}\label{eq:eps}
\kappa_{m,i}=\frac{1}{3}\times\frac{\pi^3}{30}\bar{v}_{m,i}l_{m,i}\alpha' k_B\left(\frac{k_B}{\hbar}\right)^3 T^3,
\end{equation}
respectively.

In the present work, the crystal lattice and also the magnetic structure have lower symmetry than the orthorhombic one. Using above formulas to fit the experimental data is another approximation.

\end{document}